\documentclass[aps,prd,preprint,superscriptaddress,nofootinbib]{revtex4-2}

\usepackage[T1]{fontenc}    
\usepackage[utf8]{inputenc} 
\usepackage{lmodern}        

\usepackage{amsmath,amssymb,amsthm,amsfonts} 
\usepackage{bm}             
\usepackage{slashed}        
\usepackage{physics}        

\usepackage{graphicx}       
\usepackage[caption=false]{subfig} 
\usepackage{float}          
\usepackage{booktabs}       
\usepackage{multirow}       
\usepackage{orcidlink}

\usepackage{color}
\usepackage{hyperref}       
\hypersetup{
    colorlinks=true,
    linkcolor=blue,
    citecolor=blue,
    urlcolor=blue
}

\allowdisplaybreaks

\newcommand{\Ds}{D_{s1}(2536)}
\newcommand{\dsd}{D_{s1}(2536) \to D^+_s\pi^+\pi^-}
\newcommand{\pipi}{\pi^+\pi^-}
\newcommand{\dspip}{D^+_s\pi^{+}}

\newcommand{\dspi}{D^+_s\pi^{\pm}}

\begin{document}


\preprint{}

\title{\normalsize On the \texorpdfstring{$\pi^+\pi^-$}{pi pi} and 
\texorpdfstring{$D^+_s\pi^{\pm}$}{Ds pi} mass spectra in the 
\texorpdfstring{$\dsd$}\, decays}

\author{Jorgivan Morais Dias\orcidlink{0000-0002-0354-4711}}
\email{jorgivan.dias@ufpi.edu.br}
\affiliation{\small Departamento de F\'isica, Universidade Federal do Piau\'i, 64049-550 Teresina, Piau\'i, Brasil}

\author{Yi-Yao Li\orcidlink{0009-0001-6943-4646}}
\email{liyiyao@m.scnu.edu.cn}
\affiliation{
State Key Laboratory of Nuclear Physics and 
Technology, Institute of Quantum Matter, South China Normal 
University, Guangzhou 510006, China}
\affiliation{Key Laboratory of Atomic and Subatomic Structure and Quantum Control (MOE), Guangdong-Hong Kong Joint Laboratory of Quantum Matter, Guangzhou 510006, China }
\affiliation{ Guangdong Basic Research Center of Excellence for Structure and Fundamental Interactions of Matter, Guangdong Provincial Key Laboratory of Nuclear Science, Guangzhou 510006, China  }
\affiliation{Departamento de Física Teórica and IFIC, Centro Mixto Universidad de Valencia-CSIC Institutos de Investigación de Paterna, 46071 Valencia, Spain}

\author{ Eulogio Oset\,\orcidlink{ https://orcid.org/0000-0002-4462-7919}}
\email{eulogio.oset@ific.uv.es}
\affiliation{Departamento de Física Teórica and IFIC, Centro Mixto Universidad de Valencia-CSIC Institutos de Investigación de Paterna, 46071 Valencia, Spain}
\affiliation{\small Department of Physics, Guangxi Normal University, Guilin 541004, China}

\begin{abstract}
We have carried out an evaluation of the $\pi^+ \pi^-$ and $D_s^+ \pi^+$ mass distributions in the $D_{s1}(2536)$ decay to $D_s^+ \pi^+ \pi^-$, from the perspective that the $D_{s1}(2536)$ is a molecular state, mostly made from $K^*D$ in $I=0$.  We are able to obtain, not only the mass distributions, but the branching ratio of this decay. The shape of the mass distributions differ appreciably from those of the analogous reaction $D_{s1}(2460)\to D_s^+ \pi^+ \pi^-$, which has been measured by the LHCb collaboration and analyzed theoretically from the perspective that the $D_{s1}(2460)$ is a molecular state of $D^*K$, showing a good agreement with the data. In spite of the analogy with the $D_{s1}(2460)$ decay, the dynamical differences in the decay mechanism are important, since now the $f_0(500)$ resonance is not generated, while it was the dominant mechanism in the $D_{s1}(2460)\to D_s^+ \pi^+ \pi^-$ decay. Nonetheless, we find striking differences in the mass distributions compared with phase space as a consequence of the decay mechanism. The branching ratio obtained is an order of magnitude bigger than the one of the $D_{s1}(2460)\to D_s^+ \pi^+ \pi^-$ reaction, mostly due to the larger available phase space. We also show that the shape of the distributions obtained from the molecular picture are quite different from those obtained based on a $q\bar{q}$ picture. We conclude that measuring the shape of the mass distributions and the total strength of the decay mode, should be very valuable to learn about the structure of the $D_{s1}(2536)$.
  
\end{abstract}

\maketitle

\section{Introduction}
\label{Sec:intro}

Over the past two decades, numerous new 
hadronic states have been observed, many 
of which do not fit into the conventional 
quark configurations predicted by the quark 
model, that is, $q\bar{q}$ for mesons, and 
$qqq$ for baryons 
\cite{Chen:2016qju,Lebed:2016hpi,Esposito:2016noz,Brambilla:2019esw,Meng:2022ozq}. 
While these discoveries 
challenge our understanding of the strong 
interaction, they also provide a natural 
laboratory to explore the complexity and 
richness of Quantum Chromodynamics (QCD) 
in the low-energy regime \cite{Guo:2017jvc,Albuquerque:2018jkn}.

Most of the experimental observations, as 
well as the corresponding theoretical studies, 
have focused on hadrons in the heavy-quark 
sector, involving either charm or bottom. In 
contrast, systems containing both charm and 
strangeness remain relatively unexplored, 
despite the existence of some experimental 
data available for quite some time—as is the 
case of the $D_{s1}(2536)$ resonance, with 
quantum numbers $J^P = 1^+$.

In particular, the $D_{s1}(2536)$ state 
was first observed in 1989 by the ARGUS 
Collaboration in $e^+e^-$ collisions at the 
DESY storage ring \cite{ARGUS:1989zue}. 
Its mass was measured as 
$(2535.9 \pm 0.6 \pm 2.0)$ MeV, with a narrow 
width of less than $4.6$ MeV. Since then, this 
state has been confirmed by other collaborations, 
including CLEO \cite{CLEO:1993nxj}, 
BaBar \cite{BaBar:2007cmo,BaBar:2011vbs}, 
Belle \cite{Belle:2019qoi}, and more 
recently LHCb \cite{LHCb:2023eig}. 

There have been some attempts to 
describe not only the structure of the 
$\Ds$ meson, but also to establish a 
broader understanding of the spectroscopy 
of mesons containing both charm and 
strangeness — namely, the $D_s$ mesons 
\cite{Dougall:2003hv,Dai:2003yg,Ebert:1997nk,Godfrey:1985xj,Kalashnikova:2001ig}. 
Specifically, the $\Ds$ has quantum numbers 
$J^P = 1^+$ and is therefore an axial-vector 
resonance \cite{Belle:2007kff,CLEO:1998pnu,BaBar:2011vbs}. 
In principle, this structure is 
interpreted as a quark configuration 
of the type $c\bar{s}$. However, mass 
predictions from potential models based 
on HQET \cite{Cahn:2003cw} are in disagreement 
with experimental data. This discrepancy 
also extends to other $D_{s1}$ structures, 
such as the $D_{s1}(2460)$ \cite{Ebert:2009ua}. 
In particular, the 
predicted values from quark models 
\cite{Godfrey:1985xj} lie well 
above the $D^*K$ and $K^*D$ thresholds, 
corresponding to the $D_{s1}(2460)$ and 
the $D_{s1}(2536)$, respectively. On the 
other hand, hadronic structures near 
meson-meson thresholds can be well 
described as molecular-type states 
using the molecular model \cite{Guo:2017jvc}. 
In Refs.~\cite{Kolomeitsev:2003ac,Gamermann:2007fi,Guo:2006fu,Guo:2006rp,Kong:2021ohg}, 
both the $D_{s1}(2460)$ and the $\Ds$ are 
interpreted as dynamically generated 
states arising from the interactions 
between pseudoscalar and vector mesons, 
with the $D^*K$ and $K^*D$ channels 
playing a dominant role in the formation 
of the $D_{s1}(2460)$ and $\Ds$, respectively. {The $D_{s1}(2536)$ might be different since its mass lies closer to conventional $c\bar{s}$ quark models \cite{Yang:2021tvc}. Yet, from the perspective of the molecular nature of the $D_{s1}(2460)$ and the similar dynamics that goes into the $D^*K$ and $K^*D$ interactions, one also expects a sizeable $K^*D$ component for the $D_{s1}(2536)$. The present calculation assumes the state as pure molecular, but we discuss how the experiment could eventually tell us about the size of this molecular component of the $D_{s1}(2536)$ state.}

The molecular interpretation finds 
strong support in the successful 
description of the hadronic properties 
of several resonances, both in the heavy-quark 
sector and in the light sector 
\cite{Aceti:2014uea,Dias:2014pva,Aceti:2014kja,Oller:1997ti,Roca:2005nm,Kaiser:1995eg,Oset:1997it,Oller:2000fj,Jido:2003cb,Hyodo:2008xr,Hyodo:2011qc,Sekihara:2014kya,Dong:2021bvy,Dong:2021rpi,Dong:2021juy,Guo:2013sya,Garcia-Recio:2003ejq,Wang:2019spc}. As a 
result, this framework has been extensively 
explored, with many studies proposing 
reactions in which such properties can 
be measured, putting the molecular 
picture to the test \cite{Oset:2016lyh}. 
This may shed 
light on the current debate surrounding 
the interpretation of these states as 
more complex structures, often referred 
to by the term exotic.

From this perspective, the reaction 
$D_{s1}(2460) \to D_s^+ \pi^+ \pi^-$ was 
studied in Ref.~\cite{Tang:2023yls}, assuming a molecular 
$D^*K$ interpretation for the $D_{s1}(2460)$ 
meson, and also in Ref.~\cite{Roca:2025lij} including contributions 
from the $D_s \eta$ channel, 
and using a different formalism. In this approach, 
the decay proceeds via triangle loop diagrams. 
In particular, the configuration of the 
intermediate states in the loop allows 
for the manifestation of resonances such 
as the $f_0(500)$ and even the $f_0(980)$ 
in the unitarized amplitudes associated 
with the interactions between pseudoscalar 
mesons \cite{Oller:1997ti,Kaiser:1998fi,Nieves:1998hp}. A key point in favor of this 
mechanism, as applied to the reaction 
discussed in Ref.~\cite{Tang:2023yls,Roca:2025lij}, is its good 
agreement with the corresponding 
experimental data reported by the 
LHCb Collaboration for the decay 
$D_{s1}(2460) \to D_s^+ \pi^+ \pi^-$ \cite{LHCb:2024iuo}. 
Indeed, such agreement supports the molecular 
interpretation of the $D_{s1}(2460)$ meson.

Following this line of thought, we 
assume in the present work that the 
$D_{s1}(2536)$ structure is also 
interpreted as a molecular state, dynamically 
generated by the $K^*D$ interaction, 
as suggested in Ref.~\cite{Gamermann:2007fi}. 
In order to test its molecular nature, 
we compute the invariant mass spectra 
of the $D_s^+ \pi^{\pm}$ and $\pi^+ \pi^-$ 
pairs, which are among the final products 
of the decay $D_{s1}(2536) \to D_s^+\pi^+ \pi^-$. 
In particular, we adopt the mechanism in 
which the $D_{s1}(2536)$ meson decays 
into $D_s^+ \pi^+ \pi^-$ via a triangular loop, 
with $D$, $K^*$, and $K$ mesons as intermediate 
states, in analogy to the successful mechanism employed in Ref.~\cite{Tang:2023yls,Roca:2025lij}. Through the interactions between the 
$D$ and $K$ mesons, the final states 
$D_s^+ \pi^+$ and $\pi^+ \pi^-$ are produced, 
enabling the calculation of the relevant 
spectra. {Assuming that the $D_{s1}(2536)$ state is pure molecular, this mechanism is rather predictive. The couplings needed stem from the molecular hypothesis, as well as a momentum cut off factor which enters the loop, tied to the same $K^*D$ interaction \cite{Gamermann:2007fi}. It contains an extra cut off, tied to the $KD \to D^+_s\pi^+$ transition, which is governed by the range of the light vector mesons exchanged to obtain this transition, and we quantify uncertainties from this source.} In Ref.~\cite{Roca:2025lij} the absolute rate of $D_{s1}(2460) \to D_s^+ \pi^+ \pi^-$ was also evaluated and found compatible with experiment within errors. This gives us confidence in the predictions for the rate of the $D_{s1}(2536) \to D_s^+ \pi^+ \pi^-$ decay, {yet, accepting that the $D_{s1}(2536)$ state could have a large non-molecular component, as claimed in \cite{Yang:2021tvc}, we suggest that the measurement of the absolute rate for this reaction could give us the molecular component of the state.}

This work is organized as follows: in 
Section~\ref{Sec:Formalism}, we present 
in detail the formalism employed to evaluate 
the decay $\dsd$. In particular, we describe 
the mechanism through which the reaction 
proceeds, and introduce the relevant triangle 
diagrams that contribute to this reaction. 
In the same section, we explicitly 
evaluate the triangle amplitudes 
associated with each diagram. Next, 
in Section~\ref{Sec:NumRes}, we 
present the numerical results for 
the invariant mass spectra of the $\pipi$ 
and $\dspi$ pairs, discussing in 
detail the physical implications of 
the corresponding lineshapes. Moreover, 
we compute the decay width of the 
$D_{s1}(2536)$ into the $D_s^+ \pi^+ \pi^-$ 
channel without the use of free parameters. {We devote two sections to numerical improvements and a critical discussion of the results.}
Finally, in Section~\ref{Sec:conc}, we 
summarize our main results and present our conclusions.

\section{Formalism}
\label{Sec:Formalism}

\begin{figure}[h!]  
    \centering
    \includegraphics[width=0.75\textwidth]{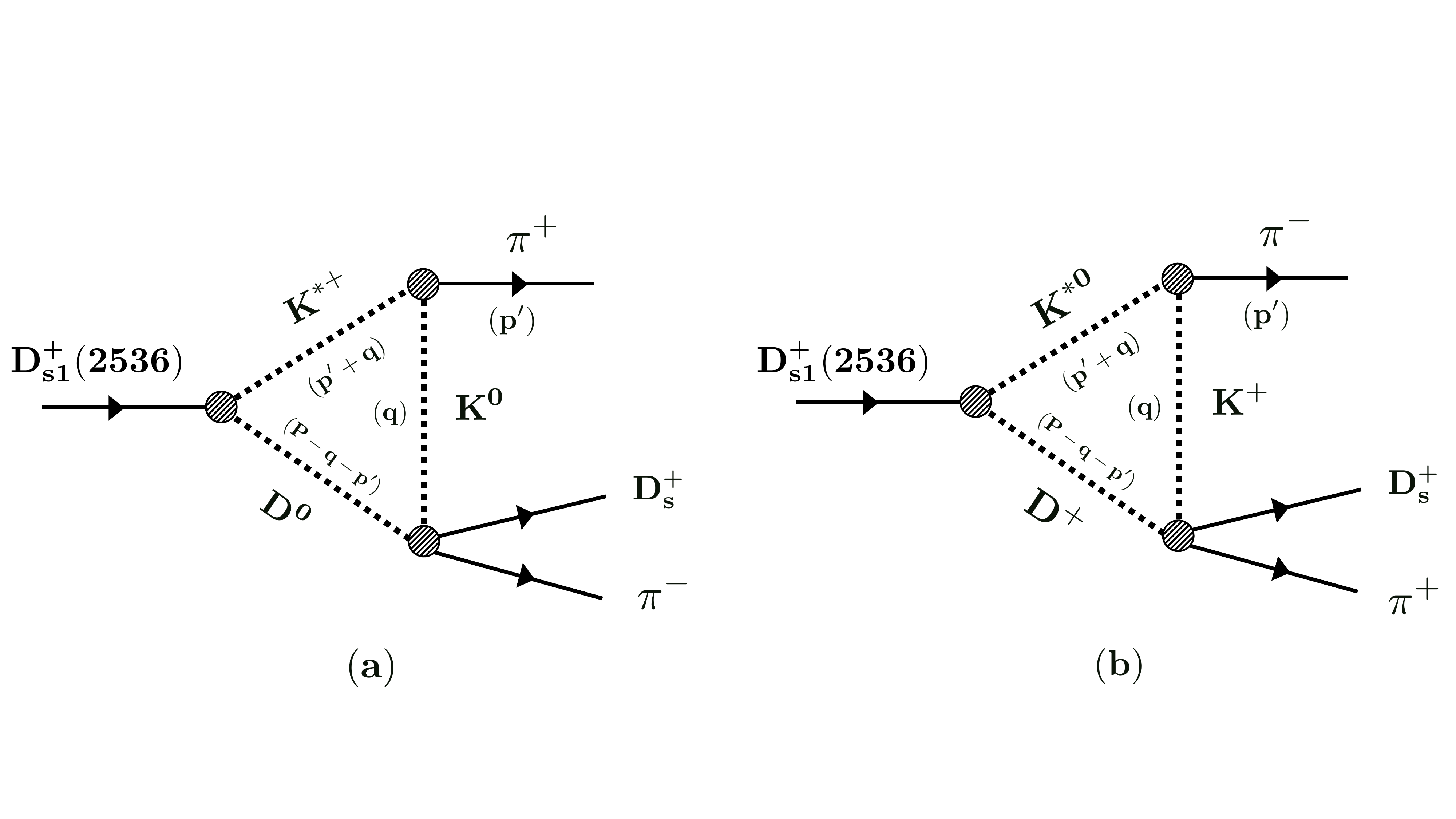} 
    \caption{$D_{s1}$ decay mechanism: (a) Decay through the 
    $K^{*+}D^0$ component, while in (b), the $D_{s1}$ meson decays through 
    the $K^{*0}D^+$ component.}
    \label{fig:ds1_mechanism}
\end{figure}

In our model, the decay $\dsd$ occurs via 
triangle loops, illustrated in Fig.~\ref{fig:ds1_mechanism}. 
In particular, we assume that the meson 
$\Ds$ is a molecular state $K^*D$, 
whose coupling to this channel is given 
in Ref.~\cite{Lin:2024hys}. In 
Fig.~\ref{fig:ds1_mechanism}, 
as a further step in the process, the meson 
$K^*$ decays into a pair of pseudoscalar mesons, 
namely: $K$ and $\pi$. The meson $K$, thus 
produced, interacts with the meson $D$ originally 
generated at the vertex $D_{s1} D K^*$, 
and the $DK$ pair undergoes a transition 
to the final $D_s \pi$ meson state, whose distribution 
interests us. The $D_{s1}(2536)$ has $I=0$. With {the} isospin phase convention 
\begin{equation}
K = \left\{
\begin{array}{ll}
K^{*+}\\
K^{*0}
\end{array} \right\} ,\,\,\,
D = \left\{
\begin{array}{ll}
\,\,D^+\\
-D^0
\end{array} \right\} ,
\end{equation}
the $K^*D$ wave function of the $D_{s1}(2536)$ is $|K^*D,I=0\rangle=-\frac{1}{\sqrt{2}}(K^{*+}D^0+K^{*0}D^+)$. The coupling of $D_{s1}(2536)$ to the $|K^*D,I=0\rangle$ state is $g_{D_{s1}(2536),K^*D,\,I=0}=20234\,\mathrm{MeV}$~\cite{Lin:2024hys}, and then $g_{D_{s1}(2536),K^{*+}D^0}=g_{D_{s1}(2536),K^{*0}D^+}=-\frac{1}{\sqrt{2}}\,g_{D_{s1}(2536),K^*D,\,I=0}$. {The vertex of $D_{s1}(2536) \to K^{*+}D^0$ corresponding to Fig.~\ref{fig:ds1_mechanism}(a) is given by}
\begin{equation}
V_{D_{s1}(2536) \to K^{*+}D^0} =g_{D_{s1}(2536),K^{*+}D^0}\,\vec{\varepsilon}_{A}\, \vec{\varepsilon}_{K^{*}}  \,,
\label{eq:V}
\end{equation}
with $\vec{\varepsilon}_{A}$ the polarization vector for the 
axial meson state $D_{s1}(2536)$. {We have neglected the $\varepsilon^0$ component of the $K^*$, but we will show in Section~\ref{Sec:consi} that this is a very good approximation.}

To calculate the amplitudes in Fig.~\ref{fig:ds1_mechanism}, 
we first need to compute the upper vertices of the $VPP$ 
type ($V$ for vector and $P$ for pseudoscalar mesons). This can be easily obtained with the help 
of the effective Lagrangians of the Local Hidden Gauge 
approach \cite{Bando:1984ej,Bando:1987br,Meissner:1987ge,Nagahiro:2008cv}. Specifically, for the $VPP$ vertices, 
we have
\begin{equation}
\mathcal{L} = -i\,g\,\langle\,\,[P,
\partial_{\mu}P\,]\,V^{\mu}\,\rangle\, ,
\end{equation}
where $g=M_V/2f$, with $M_V=800$ MeV, an average vector mesons mass, while $f$ 
is the pion decay constant, equal to 93 MeV, and the symbol $\left<...\right>$ denotes the trace of the SU(3) matrices. 
Additionally, $P$ and $V^{\mu}$ correspond to 
the matrices $q_i\bar{q}_j$ in terms of quarks, given in terms of the physical pseudoscalar and vector mesons, respectively, by

\begin{align} & P=\left(\begin{array}{cccc}\frac{\pi^0}{\sqrt{2}}+\frac{\eta}{\sqrt{3}} 
    & \pi^{+} & K^{+} & \bar{D}^0 \\ \pi^{-} & -\frac{\pi^0}{\sqrt{2}}+\frac{\eta}{\sqrt{3}} 
    & K^0 & D^{-} \\ K^{-} & \bar{K}^0 & -\frac{\eta}{\sqrt{3}} & D_s^{-} \\ D^0 & D^{+} 
    & D_s^{+} & \eta_c\end{array}\right)\,, \\ & V=\left(\begin{array}{cccc}\frac{\rho^0}{\sqrt{2}}
        +\frac{\omega}{\sqrt{2}} & \rho^{+} & K^{*+} & \bar{D}^{* 0} \\ 
        \rho^{*-} & -\frac{\rho^0}{\sqrt{2}}+\frac{\omega}{\sqrt{2}} & K^{* 0} 
        & D^{*-} \\ K^{*-} & \bar{K}^{* 0} & \phi & D_s^{*-} \\ D^{* 0} & D^{*+} 
        & D_s^{*+} & J / \psi\end{array}\right),
\end{align}
where we take into account the standard 
$\eta-\eta^{\prime}$ mixing, as discussed in 
Ref.~\cite{Bramon:1992kr}. Thus, considering the 
momenta defined in Fig.~\ref{fig:ds1_mechanism}, 
we obtain the following expression 
for the relevant $VPP$ vertices
\begin{equation}
V_{VPP} = g\, \varepsilon_{\mu}\,(q - p^{\prime})^{\mu}\, ,
\end{equation}
where $\varepsilon_{\mu}$ is the polarization 
vector of the meson $K^*$.

Next, we need to calculate the transitions 
$D^0 K^0 \to D_s^+ \pi^-$,$D^+ K^+ \to D_s^+ \pi^+$, 
which appear at the lower vertices of the diagrams 
in Fig.~\ref{fig:ds1_mechanism}. These $PP\to PP$ 
transitions (with $P$ standing for pseudoscalar meson) 
can be easily computed using an extension of the 
Local Hidden Gauge approach \cite{Bando:1984ej,Bando:1987br,Meissner:1987ge,Nagahiro:2008cv}. 
They were calculated in Ref.~\cite{Roca:2025lij}.  
Specifically, for the $D^0 K^0 \to D_s^+ \pi^-$ we have
\begin{equation}
    V_{K^0D^0\to D^+_s\pi^-} = \frac{1}{4f^2}\,\frac{1}{2}\,\Big[ 
        3s - (m^2_{K^0} + m^2_{\pi^-} + m^2_{D^0} + m^2_{D_s^+}) - \frac{1}{s}
        \,(m^2_{D^0} - m^2_{K^0})\,(m^2_{D_s^+} - m^2_{\pi^-})\Big]\, ,
        \label{eq:v_dk}
\end{equation}
where $s \equiv M_{\mathrm{inv}}^2(D_s^+\pi^-)$.

Due to isospin symmetry, the transition potentials 
for the processes $D^0 K^0 \to  D_s^+\pi^-$ and 
$D^+ K^+ \to  D_s^+\pi^+$ are identical in both 
magnitude and sign. This follows from the 
fact that both initial two-meson states 
couple to a total isospin $I=1$, differing 
only in their third component $I_3$. 
Furthermore, by carefully considering 
the conventional phase assignments of 
the isospin multiplets—particularly the 
$(- \pi^+, \pi^0, \pi^-)$ basis for pions 
and the $D,K$ isospin phase conventions given above—the relative phases between the 
physical states ensure that no additional 
sign difference arises between these transitions. 
Consequently, the interaction potential 
derived for the $D^0 K^0 \to  D_s^+\pi^-$ channel 
in Eq.~\eqref{eq:v_dk} can be directly applied to the 
$D^+ K^+ \to  D_s^+\pi^+$ channel.

Once the relevant vertices for calculating 
the diagrams in Fig.~\ref{fig:ds1_mechanism}(a)-(b) 
have been obtained, we can write the corresponding 
amplitudes for the triangle loops under consideration. 
Next, we will discuss in detail the loop in 
Fig.~\ref{fig:ds1_mechanism}(a), while the 
calculation for the diagram in Fig.~\ref{fig:ds1_mechanism}(b) 
follows analogously. With the upper vertices $K^{*+} \pi^{+} K^0$ 
and $K^{*0} \pi^{-} K^{+}$ at hand, and considering 
Eq.~\eqref{eq:v_dk} for the transition $D^0 K^0 \to  D_s^{+}\pi^{-}$, 
the diagram in Fig.~\ref{fig:ds1_mechanism}(a) can be written as
\begin{align}
    & -i t^{(a)}_{\textrm{loop}} = \int \frac{d^4 q}{(2 \pi)^4} (-i)g_{D_{s1}(2536),K^{*+} D^0}\,
    {\varepsilon}_{Ai}\, {\varepsilon}_{K^*i}\, \frac{i}{\left(p^{\prime}
    +q\right)^2-m_{K^*}^2+i \varepsilon}(-i)\, g\,\,
    \varepsilon_{K^{*}\alpha}\,\left(q-p^{\prime}\right)^{\alpha} \nonumber\\ 
    & \times\frac{i}{q^2-m_K^2+i \varepsilon}\,
    \frac{i}{\left(P-p^{\prime}-q\right)^2-m_D^2+i \varepsilon}(-i)
    V_{D^0K^0\to D^+_s\pi^-}\, ,
    \label{eq:def_loop}
\end{align}
with $P$ standing for the momentum of the $D_{s1}(2536)$.

Assuming we are dealing with small three-momenta 
compared to the masses of the particles involved, it implies 
that only the spatial components of the polarization vector of 
$K^*$ meson are non-vanishing, that is
\begin{equation}
    \varepsilon^0_{K^*} = \frac{\left|\,\vec{p^{\,\prime}}+\vec{q}\,\right|}{m_{K^*}} 
    \simeq 0\, ,
    \label{eq:appxi}
\end{equation}
and the completeness relation for the corresponding 
polarization vectors, which reads
\begin{equation}
    \sum\limits_{\textrm{pol}} \, \varepsilon_{K^*\mu}\,\varepsilon_{K^*\alpha}\,=
    -g_{\mu\alpha} + \frac{(p^{\prime}+ q)_{\mu}\,(p^{\prime}+ q)_{\alpha}}{m^2_{K^*}}\, ,
\end{equation}
can now be written as
\begin{equation}
    \sum\limits_{\textrm{pol}} \varepsilon_{K^*\mu}\,\varepsilon_{K^*\alpha} 
    \simeq \sum\limits_{\textrm{pol}} \varepsilon_{K^*i} \varepsilon_{K^*j} \simeq 
    \delta_{ij}\, .
    \label{eq:pol_vec}
\end{equation}
{In Section~\ref{Sec:consi} we will actually see that the approximation of Eq.~\eqref{eq:appxi} is indeed very good.}

Therefore, using Eq.~\eqref{eq:pol_vec}, the triangle amplitude reduces to
\begin{align} & -i t^{(a)}_{\textrm{loop}} = \int \frac{d^4 q}{(2 \pi)^4}(-i) g_{D_{s1}(2536),K^{*+} D^0}\,
    \varepsilon_{Ai} \varepsilon_{K^{*}i}\, \frac{i}{\left(p^{\prime}
    +q\right)^2-m_{K^*}^2+i \varepsilon}(-i)\, g\,
    \varepsilon_{K^* j} \nonumber\\ 
    & \times\left(p^{\prime}-q\right)_j 
    \,\,\frac{i}{q^2-m_K^2+i \varepsilon}\,
    \frac{i}{\left(P-p^{\prime}-q\right)^2-m_D^2+i \varepsilon}(-i)
    V_{D^0K^0\to D^+_s\pi^-}\, .
    \label{eq:loop_def}
\end{align}

{For the heavy mesons $K^{*+}$, $D^0$, coming from the decay of a particle, it is a good approximation to take the positive energy part of their full propagators,}  that is
\begin{align}
    \frac{1}{(p^{\prime}+ q)^2 - m^2_{K^*}+i\,\varepsilon} \simeq &
    \,\frac{1}{2\,\omega_{K^*}(\vec{p}^{\,\prime}+\vec{q}\,)} \frac{1}{p^{\prime\,0} + q^0 - 
    \omega_{K^*}(\vec{p}^{\,\prime}+\vec{q}\,)+i\varepsilon} \,,\nonumber\\
    \frac{1}{\left(P-q-p^{\prime}\right)^2-m_D^2+i \varepsilon} & \simeq 
    \frac{1}{2\omega_{D}(\vec{p}^{\,\prime}+\vec{q}\,)} \frac{1}{P^0-p^{\prime\,0}-q^0 - 
    \omega_D(\vec{p}^{\,\prime}+\vec{q}\,)+i\varepsilon}\, ,
\end{align}
where $\omega_{K^*}(\vec{p}^{\,\prime}+\vec{q}\,) = \sqrt{(\vec{p}^{\,\prime}+\vec{q}\,)^2 + m^2_{K^*}}$ ,  
$\omega_D(\vec{p}^{\,\prime}+\vec{q}\,) = \sqrt{(\vec{p}^{\,\prime}+\vec{q}\,)^2 + m^2_D}$ , $\omega_K(\vec{q}\,) = \sqrt{\vec{q}\,^2 + m^2_K}$ (for later use), and $\vec{P}=0$ in the $D_{s1}$ rest frame.
On the other hand, for the $K$ meson, we keep 
both positive and negative energy parts 
since it is a lighter meson and can be more off shell. {In Section~\ref{Sec:consi} we shall see the goodness of the approximation.} Hence, Eq.~\eqref{eq:loop_def} 
can be rewritten as 
\begin{align}
    &  t^{(a)}_{\textrm{loop}} = i\,g\,g_{D_{s1}(2536),K^{*+} D^0}\,
    \,V_{D^0K^0\to  D^+_s\pi^-}\,\varepsilon_{Ai}\,\int \frac{d^4 q}{(2 \pi)^4}\,
    (p^{\prime}- q)_i\,\frac{1}{2\omega_{K^*}}\,\frac{1}{2\omega_K}\,\frac{1}{2\omega_D}\nonumber\\
    & \times \frac{1}{p^{\prime 0}+ q^0 - \omega_{K^*}(\vec{p}^{\,\,\prime}+\vec{q}\,) 
    + i\varepsilon}\,\frac{1}{P^0-p^{\prime 0}-q^0 - \omega_D(\vec{p}^{\,\,\prime}+\vec{q}\,) + 
    i\varepsilon}\nonumber\\
    &\times \Big(\frac{1}{q^0 - \omega_K(\vec{q}\,)+ i\varepsilon} - 
    \frac{1}{q^0 + \omega_K(\vec{q}\,) - i\varepsilon}\Big)\, .
    \label{eq:loop_def3}
\end{align}
Now, we can perform an analytical integration over the $q^0$ 
variable, using Cauchy's residue theorem, so that Eq.~\eqref{eq:loop_def3} 
now assumes the following form
\begin{align}
    &  t^{(a)}_{\textrm{loop}} = g\,g_{D_{s1}(2536),K^{*+} D^0}\,
    \,V_{D^0K^0\to \pi^- D^+_s}\,\,\vec{\varepsilon}_{Ai}\,\int \frac{d^3 q}{(2 \pi)^3}\,
    (p^{\prime}- q)_i\,\frac{1}{2\omega_{K^*}}\,\frac{1}{2\omega_K}\,\frac{1}{2\omega_D}\,
    \nonumber\\
    & \times \frac{1}{P^0 - \omega_D(\vec{p}^{\,\,\prime}+\vec{q}\,) - \omega_{K^*}(\vec{p}^{\,\,\prime}+\vec{q}\,) 
    + i\varepsilon}\,\Big(\frac{1}{P^0-p^{\,\prime 0} - \omega_D(\vec{p}^{\,\,\prime}+\vec{q}\,) - 
    \omega_K(\vec{q}\,) + i\varepsilon} \nonumber \\
    &+ \frac{1}{p^{\prime \,0} - 
    \omega_{K^*}(\vec{p}^{\,\,\prime}+\vec{q}\,) - \omega_K(\vec{q}\,)+i\varepsilon} \Big)\, .
    \label{eq:loop_def4}
\end{align}

Next, considering that 
\begin{equation}
    \int\, d^3q\, F(\vec{q}, \vec{p}^{\,\,\prime})\,q_{\ell} = p^{\prime}_{\ell}\,\,
    \int\, d^3q\, F(\vec{q}, \vec{p}^{\,\,\prime})\,\,
    \frac{\vec{q}\cdot \vec{p}^{\,\,\prime}}{|\vec{p}^{\,\,\prime\,}|^2}\, ,
\end{equation}
we finally write Eq.~\eqref{eq:loop_def4} as
\begin{equation}
    t^{(a)}_{\textrm{loop}} = \mathcal{I}^{(a)}_{loop}\,\,\left(\vec{\varepsilon}_A\cdot \vec{p}^{\,\,\prime}\right)\,,
    \label{eq:loop_1a}
\end{equation}
with $\mathcal{I}^{(a)}_{loop}$ given by
\begin{align}
    &\mathcal{I}^{(a)}_{loop} = g\,g_{D_{s1}(2536),K^{*+} D^0}\,\,V_{D^0K^0\to  D^+_s\pi^-}\,
    \int\,\frac{d^3q}{(2\pi)^3}\,\Theta\left(q_{\max}-
    \left|\vec{q}+\vec{p}^{\,
\prime}\right|\right)\,\Theta\left(\Lambda-|q^*|\right)\left( 1 - 
    \frac{\vec{q}\cdot \vec{p}^{\,\,\prime}}{|\vec{p}^{\,\,\prime}|^2} \right)\,
    \nonumber\\
    &\times \,\frac{1}{2\omega_{K^*}(\vec{p}^{\,\,\prime}+\vec{q}\,)}\frac{1}{2\omega_D(\vec{p}^{\,\,\prime}+\vec{q}
    \,)}\,
    \frac{1}{2\omega_K(\vec{q}\,)}\frac{1}{P^0-
    \omega_D\left(\vec{p}^{\,\prime}+\vec{q}\,\right)-\omega_{K^*}\left(\vec{p}^{\,\prime}
    +\vec{q}\,\right) + i\frac{\Gamma_{K^*}}{2}}\nonumber\\
    &\times \,\Big( \frac{1}{P^0 - p^{\prime\,0} -
    \omega_{D}(\vec{p}^{\,\,\prime}+\vec{q}\,)-\omega_K(\vec{q}\,)+ i\varepsilon} + \frac{1}{p^{\prime 0} 
    - \omega_{K^*}(\vec{p}^{\,\,\prime}+\vec{q}\,) - \omega_K(\vec{q}\,)+i\frac{\Gamma_{K^*}}{2}} \Big)\, ,
    \label{eq:int_loop}
\end{align}
where $\vec{p}^{\,\,\prime}$ in this case, corresponds 
to the $\pi^+$ momentum, namely,
\begin{equation}
    \vec{p}^{\,\,\prime} = \frac{\lambda^{\frac{1}{2}}\left(m^2_{D_{s1}}, m^2_{\pi^+}, M^2_{\mathrm{inv}}(D_s^+\pi^-)\right)}{2\,m_{D_{s1}}}\, .
\end{equation}

In Eq.~\eqref{eq:int_loop} we have implemented two changes, the introduction of {two} $\Theta$ functions and the explicit consideration of the $K^*$ width. The appearance in Eq.~\eqref{eq:int_loop} 
of a  Heaviside theta function 
$\Theta$ which sets the upper limit in the 
integral, comes naturally from the chiral 
unitary approach, which considers a separable 
potential of the form $V(\vec{q}^{\,\prime}, 
\vec{q}) = V \Theta\left(q_{\textrm{max}}-
|\vec{q}^{\,\prime}\,|\right)\, 
\Theta\left(q_{\textrm{max}}-|\vec{q}\,|\right)$, 
leading to $T(\vec{q}^{\,\prime}, \vec{q}\,) = 
T\, \Theta\left(q_{\textrm{max}}-
\left|\vec{q}^{\,\prime}\right|\right) 
\Theta\left(q_{\textrm{max}}-|\vec{q}\,
|\right)$, where $q_{\textrm{max}}$ stands 
for the cutoff used to regularize the loops 
encoded in the $G$-function. In Ref.~\cite{Lin:2024hys}, 
the $D_{s1}(2536)$ state is dynamically 
generated from $K^*D$ interaction with 
$q_{\textrm{max}}=1025$ MeV, which is the 
value we are considering in Eq.~\eqref{eq:int_loop}. {Another theta function $\Theta\left(\Lambda-|q^*|\right)\ $ is for the vertex $D^0K^0 \to D_s^+\pi^-$, where $q^*$ is the three momentum of $K^0$ boosted to $D_s^+\pi^-$ rest frame}
\begin{equation}
    q^* = \left[ \left(\frac{E_{D_s^+\pi^-}}{M_{\mathrm{inv}}(D_s^+\pi^-)}-1 \right)\frac{\vec{q}\cdot\vec{p}^{\,\prime} }{\vec{p}^{\,\prime2}}+\frac{E_{K^0}}{M_{\mathrm{inv}}(D_s^+\pi^-)}\right]\cdot\vec{p}^{\,\prime}+\vec{q}\, ,
\end{equation}
{where $E_{D_s^+\pi^-}=\sqrt{M^2_{\mathrm{inv}}(D_s^+\pi^-)+|\vec{p}^{\,\prime}|^2}$ , $E_{K^0}=\sqrt{m^2_{K^0}+|\vec{q}\,|^2}$ . We calculate with $\Lambda\equiv 600\,\mathrm{MeV}$ and vary it within the range $\Lambda \in \left[500,700\right]$ MeV to estimate the uncertainty in our result as done in Ref.~\cite{Roca:2025lij} (the change induced by changing $q_{\mathrm{max}}$ are very small compared to those changing $\Lambda$).} {The values of $q_{\max}$ or $\Lambda$ are not arbitrary, they are tied to the range of the interaction in momentum space, which is governed by the exchange of light vector mesons \cite{Gamermann:2009uq}. The values 500-700 MeV are common in most problems. The fact that $q_{\max}$, tied to the $K^*D$ interaction is somewhat bigger when forcing the $D_{s1}(2536)$ to be fully molecular is an indication that other components should also play a role in the building up of this state.} 

The triangle loop in Fig.~\ref{fig:ds1_mechanism}(b) 
is readily obtained from Eq.~\eqref{eq:loop_1a} by setting 
$\vec{p}^{\,\prime} \to \vec{p}^{\,\prime}_{\pi^-}$, that is,
\begin{equation}
    t^{(b)}_{\textrm{loop}} = \mathcal{I}^{(b)}_{loop}\,\,\left(\vec{\varepsilon}_A\cdot 
    \vec{p}^{\,\,\prime}_{\pi^-}\right)\,,
    \label{eq:loop_1b}
\end{equation}
where $t^{(b)}_{\textrm{loop}}$ is
\begin{align}
    &\mathcal{I}^{(b)}_{loop} = g\,g_{D_{s1}(2536),K^{*0} D^+}\,\,V_{D^+K^+\to  D^+_s\pi^+}\,
    \int\,\frac{d^3q}{(2\pi)^3}\,\Theta\left(q_{\max}-
    \left|\vec{q}+\vec{p}^{\,\prime}_{\pi^-}\right|\right)\,\Theta(\Lambda-|q^*|)\left( 1 - 
    \frac{\vec{q}\cdot \vec{p}^{\,\,\prime}_{\pi^-}}{|\vec{p}^{\,\,\prime}_{\pi^-}|^2} \right)\,
    \nonumber\\
    &\times \,\frac{1}{2\omega_{K^*}(\vec{p}^{\,\,\prime}_{\pi^-}+\vec{q}\,)}\,\frac{1}{2\omega_D(\vec{p}^{\,\,\prime}_{\pi^-}+\vec{q}\,)}\,
    \frac{1}{2\omega_K(\vec{q}\,)}\frac{1}{P^0-
    \omega_D\left(\vec{p}^{\,\prime}_{\pi^-}+\vec{q}\,\right)-\omega_{K^*}\left(\vec{p}^{\,\prime}_{\pi^-}
    +\vec{q}\,\right) + i\frac{\Gamma_{K^*}}{2}}\nonumber\\
    &\times \,\Big( \frac{1}{P^0 - p^{\prime\,0}_{\pi^-} -
    \omega_{D}(\vec{p}^{\,\,\prime}_{\pi^-}+\vec{q}\,)-\omega_K(\vec{q}\,)+ i\varepsilon} + \frac{1}{p^{\prime 0}_{\pi^-} 
    - \omega_{K^*}(\vec{p}^{\,\,\prime}_{\pi^-}+\vec{q}\,) - \omega_K(\vec{q}\,)+i\frac{\Gamma_{K^*}}{2}} \Big)\, .
    \label{eq:int_loopb}
\end{align}

As a result, the total amplitude will be 
\begin{equation}
    t_{\textrm{total}} = \mathcal{I}^{(a)}_{\textrm{loop}}\,
    (\vec{\varepsilon}_A\cdot \vec{p}^{\,\,\prime}_{\pi^+}) + 
    \mathcal{I}^{(b)}_{\textrm{loop}}\,
    (\vec{\varepsilon}_A\cdot \vec{p}^{\,\,\prime}_{\pi^-})\, ,
    \label{eq:total_amp}
\end{equation}
where $\vec{p}^{\,\,\prime}_{\pi^+}$ and $\vec{p}^{\,\,\prime}_{\pi^-}$ are 
the $\pi^+$ and $\pi^-$ momenta in the $D_{s1}$ rest frame, 
\begin{align}
    & \vec{p}^{\,\,\prime}_{\pi^+} = \frac{\lambda^{\frac{1}{2}}\left({m^2_{D_{s1}}, m^2_{\pi^+}, 
    m^2_{23}}\right)}{2\,m_{D_{s1}}}\, ,\nonumber\\
    & \vec{p}^{\,\,\prime}_{\pi^-} = \frac{\lambda^{\frac{1}{2}}\left({m^2_{D_{s1}}, m^2_{\pi^-}, 
    m^2_{13}}\right)}{2\,m_{D_{s1}}}\,,
\end{align}
with $m_{23}$ and $m_{13}$ standing for the invariant masses associated with 
the $D^+_s\pi^-$ and $D^+_s\pi^+$ pairs, respectively ($\pi^+(1)$, $\pi^-(2)$, $D_s(3)$). Furthermore, these masses are not all independent, and by using 
\begin{equation}
    m^2_{12} + m^2_{23} + m^2_{13} = m^2_{D_{s1}} + m^2_{D_s^+} + m^2_{\pi^+} + m^2_{\pi^-}\, ,
    \label{eq:inv_masses}
\end{equation}
we can evaluate Eq.~\eqref{eq:total_amp} as a function of the 
invariant masses that we are interested in.

Averaging over the $D_{s1}$ polarizations, and taking the modulus 
squared of the amplitude in Eq.~\eqref{eq:total_amp}, that is
\begin{align} 
    & \overline{\sum}\,\, \sum\left|t_{\textrm{total}}\right|^2 = \frac{1}{3}\Bigg\{\left|\mathcal{I}_{\textrm{loop}}^{(a)}\right|^2\left|\vec{p}_{\pi^{+}}^{\,\prime}\right|^2
    +\left|\mathcal{I}_{\textrm{loop}}^{(b)}\right|^2\left|\vec{p}_{\pi^-}\right|^2+ 
    2\, \textrm{Re}\left[\mathcal{I}_{\textrm{loop}}^{(a)} \cdot\left(\mathcal{I}_{\textrm{loop}}^{(b)}\right)^*\right]
    \left(\vec{p}_{\pi^{+}}^{\,\prime} \cdot \vec{p}_{\pi^{-}}^{\,\prime}\right)\Bigg\} \, ,
    \label{eq:loop_amp_tot}
\end{align}
we are able to obtain the double differential invariant mass 
distribution \cite{ParticleDataGroup:2024cfk}, which is
\begin{equation}
    \frac{d \Gamma}{d m_{12}\, d m_{23}}=\frac{1}{(2 \pi)^3} \frac{2\, m_{12}\, 2\, m_{23}}{32\, m_{D_{s 1}}^3}
    \,\overline{\sum}\,\sum\left|t_{\textrm{total}}\right|^2 \, .
    \label{eq:double_mass_spectra}
\end{equation}
Since all the momenta involved and the scalar 
products in Eq.~\eqref{eq:loop_amp_tot} are given in terms 
of the invariant masses $m_{12}$, $m_{23}$, and $m_{13}$, by using 
Eq.~\eqref{eq:inv_masses} we can obtain the two-particle 
mass distribution in terms of any one of those invariant masses. 
For instance, for the $m_{12}$ (the $\pi^+\pi^-$) distribution we 
integrate Eq.~\eqref{eq:double_mass_spectra} over $m_{23}$, such that
\begin{equation}
    \frac{d\Gamma}{dm_{12}} = \int\limits^{m^{\textrm{max}}_{23}}_{m^{\textrm{min}}_{23}}\, 
    \frac{d\Gamma}{dm_{12}\,dm_{23}}\,dm_{23}\, ,
\end{equation}
with $m^{\textrm{min}}_{23}$, and $m^{\textrm{max}}_{23}$ given by 
\begin{align}
    & m^{\textrm{min}}_{23} = \sqrt{(E_2 + E_3)^2 - (\sqrt{E_2^2 - m^2_2}+
    \sqrt{E_3^2 - m^2_3})^2}\, ,\nonumber\\
    & m^{\textrm{max}}_{23} = \sqrt{(E_2 + E_3)^2 - (\sqrt{E_2^2 - m^2_2}-
    \sqrt{E_3^2 - m^2_3})^2}\, .
\end{align}
Moreover, $E_2$, $E_3$ are given by
\begin{align}
    & E_2 = \frac{m^2_{12} - m^2_{1} + m^2_{2}}{2\,m_{12}}\,,\nonumber\\
    & E_3 = \frac{m^2_{D_{s1}} - m^2_{12} - m^2_{3}}{2\,m_{12}}\, .
\end{align}
All the other two-particle mass distribution are easily 
derived by taking cyclical permutations 
of the $1,\,2,\,3$ indices in the expressions above.

\section{Numerical Results}
\label{Sec:NumRes}

\begin{figure}[h!]  
    \centering
    \includegraphics[width=0.6\textwidth]{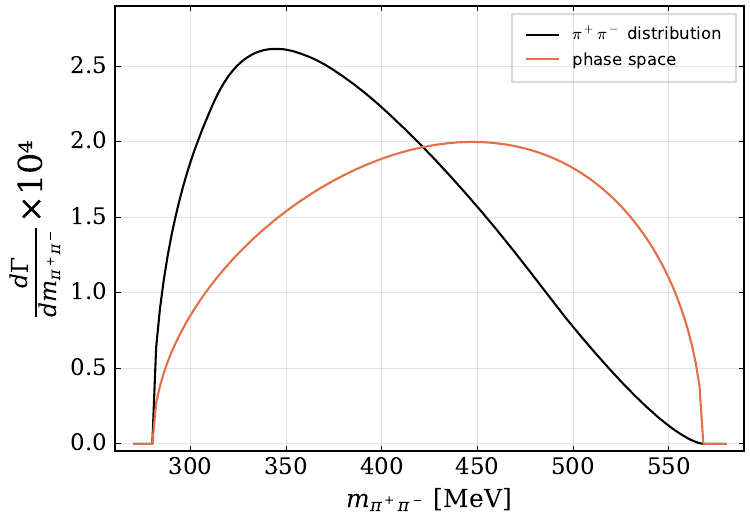} 
    \caption{$\pipi$ invariant mass spectrum and phase space.}
    \label{fig:pipi}
\end{figure}

\begin{figure}[h!]  
    \centering
    \includegraphics[width=0.6\textwidth]{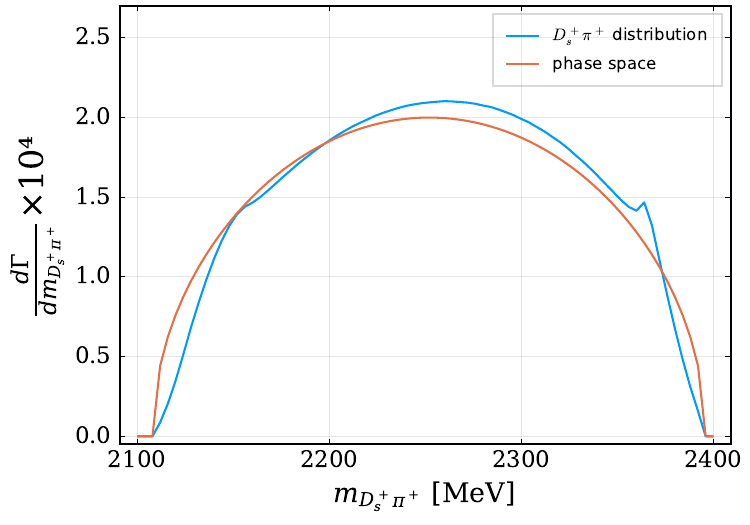} 
    \caption{$\dspip$ invariant mass spectrum and phase space. }
    \label{fig:dspi}
\end{figure}

The numerical results are presented in Figs.~\ref{fig:pipi} 
and \ref{fig:dspi}. Specifically, Fig.~\ref{fig:pipi} 
illustrates the $\pi^+ \pi^-$ mass distribution. We can 
observe a smooth lineshape throughout the invariant 
mass range allowed by kinematics, including a broad 
peak in the range of $300$ MeV to $400$ MeV. It is interesting to mention that even if we do not have a direct $\pi^+\pi^-$ interaction here, the shape of the mass distribution is quite different from one of pure phase space. This is due to the particular mechanism of $D_s^+\pi^+\pi^-$ production, tied to the molecular nature of the $D_{s1}(2536)$ and the triangle mechanism that follows as a consequence.

In Fig.~\ref{fig:dspi}, on the other hand, we 
notice a smooth curve throughout most of the 
distribution, except at the extremes, where two 
cusp structures are clearly visible: one around 
$2140$ MeV and another, more pronounced, in the 
high-energy region of the spectrum, at approximately 
$2363$ MeV. This latter value is associated with a threshold 
singularity, arising from the contribution of the 
$D^0 K^0 \to  D^+_s\pi^-$ transition exactly at 
the $D^0 K^0$ production threshold, located at $2363$ 
MeV. This threshold condition consequently generates 
a cusp at this energy.

Similarly, the small enhancement observed around 
$2140$ MeV can be understood as a result of the on-shell 
contribution of the $D^+ K^+$ pair in the $D^+ K^+ \to  D^+_s\pi^+$ 
transition. In particular, we expect the appearance 
of a cusp in the mass spectrum shown in Fig.~\ref{fig:dspi}, 
precisely at $2143$ MeV, when the invariant mass of the 
$ D^+_s\pi^+$ system, $m_{D^+_s\pi^+}$, reaches 
approximately $2363$ MeV.

This correlation between the two values can 
be understood by noting that, in the rest 
frame of the $ D^+_s\pi^+$ pair, the invariant 
mass $m_{D^+_s\pi^-}$ can be written as a function 
of $m_{D^+_s\pi^+}$ through the relation
\begin{equation}
    m^2_{ D^+_s\pi^-} = m^2_{\pi^-} + m^2_{D_s^+} + 
    2\,E_{\pi^-}\, E_{D_s^+} - 2\,\left(\vec{p}_{\pi^-} \cdot \vec{p}_{D_s^+}\right) \, ,    
\end{equation}
where $\vec{p}_{\pi^-}$ and $\vec{p}_{D_s^+}$ are the 
three-momenta of the $\pi^-$ and $D^+_s$ mesons, 
respectively, evaluated in the $ D^+_s\pi^+$ 
rest frame
\begin{align}
    & \vec{p}_{\pi^-} = \frac{\lambda^{\frac{1}{2}}\left({m^2_{D_{s1}}, m^2_{\pi^-}, 
    M^2_{\mathrm{inv}}(D_s^+\pi^+)}\right)}{2\,M_{\mathrm{inv}}(D_s^+\pi^+)}\, ,\nonumber\\
    & \vec{p}_{D_s^+} = \frac{\lambda^{\frac{1}{2}}\left({M^2_{\mathrm{inv}}(D_s^+\pi^+), m^2_{\pi^-}, 
    m^2_{D_s^+}}\right)}{2\,M_{\mathrm{inv}}(D_s^+\pi^+)}\,.
\end{align}
Both $E_{\pi^-}$, $E_{D_s^+}$, and the magnitudes 
of the momenta depend on $m_{D^+_s\pi^+}$. 
Therefore, by setting $m_{D^+_s\pi^+} = 2363$ 
MeV, the above equation yields $m_{ D^+_s\pi^-} 
\approx 2143$ MeV, precisely where the 
corresponding cusp is expected to appear. It should be noticed that in this case the phase space distribution is similar to the actual one, except for the two cusps discussed above.

We do not plot the $D_s^+\pi^-$ mass distribution which is very similar to the $D_s^+\pi^+$ one (slight difference because of  different masses of the intermediate states).

{As we have mentioned, with the assumption of $D_{s1}(2536)$ as a pure $K^*D$ molecule, we can make absolute predictions of the mass distributions and the integrated decay width, up to the dependence on $\Lambda$.} This width is obtained by determining the 
area under the spectral curves in Figs.~\ref{fig:pipi} 
and \ref{fig:dspi}, which should, of course, yield 
the same value. Consequently, we obtain the following 
value for the decay width of the $\dsd$ process
\begin{equation}
    \Gamma(\dsd) = 44\pm15\, \mathrm{keV}\, ,
    \label{eq:widthresult}
\end{equation}
{where the uncertainty arises from the variation 
of the cutoff parameter $\Lambda$.} This value is about an order of magnitude bigger than the one obtained in the $D_{s1}(2460) \to D_s^+ \pi^+ \pi^-$ decay~\cite{Roca:2025lij}, which should make its observation feasible.

\section{Consideration of $\varepsilon^0(K^*)$ and negative energy parts of the propagators}\label{Sec:consi}

\subsection{Consideration of $\varepsilon^0(K^*)$}

{We should start with a covariant $V_{D_{s1}(2536) \to K^{*+} D^0}$ vertex substituting in Eq.~\eqref{eq:V}
\begin{equation}
    \vec{\varepsilon}_A~\vec{\varepsilon}_{K^*} = -\varepsilon_{A\mu}~\varepsilon^{~\mu}_{K^*}\, .
\end{equation}
Actually, since the $D_{s1}(2536)$ state is at rest, $\varepsilon_A^{0}=0$, and we would have the vertex
\begin{equation}
    \varepsilon_{Ai}~\varepsilon_{K^*i}\, .
\end{equation}
Then we would have for the combination of Eq.~\eqref{eq:def_loop}
\begin{equation}
\sum_{K^* pol}\varepsilon_{Ai}~\varepsilon_{K^*i}~\varepsilon_{K^*\alpha}~(q-p')^\alpha=\varepsilon_{Ai}(-g_{i\alpha}+\frac{(p'+q)_i(p'+q)_\alpha}{m_{K^*}^2})(q-p')^\alpha\, .
\label{eq:epsilon}
\end{equation}
The first term in Eq.~\eqref{eq:epsilon} gives rise to what we have calculated and we get a new term
\begin{equation}
\frac{\varepsilon_{Ai}(p'+q)_i(q^2-p'^2)}{m_{K^*}^2}\, ,
\label{eq:new}
\end{equation}
which should be compared to
\begin{equation}
\varepsilon_{Ai}(q-p')_i\, .
\label{eq:old}
\end{equation}
With the value of $q^0=P^0-p'^0-\omega_D(\vec{p}~'+\vec{q}~)$ that we get from the Cauchy integration we find that $(q^2-p'^2)/m_{K^*}^2\approx0.03$. Furthermore, the term of Eq.~\eqref{eq:new} goes with $\varepsilon_{Ai}(p'+q)_i$ where $\vec{p}~'+\vec{q}$ will be zero in average, while the term in Eq.~\eqref{eq:old} goes as $(q-p')_i$, which does not vanish as an average. We can conclude that the new term is basically negligible.
}

\subsection{Consideration of negative energy component of the propagators}

{In Eq.~\eqref{eq:loop_def3} we will now consider the negative energy component of the $K^*$ propagator
\begin{equation}
-\frac{1}{2\omega_{K^*}(\vec{p}~'+\vec{q}~)}\frac{1}{p'^0+q^0+\omega_{K^*}(\vec{p}~'+\vec{q}~)-i\varepsilon}\, 
\end{equation}
and we apply again Cauchy's theorem to evaluate the term. We obtain in Eq.~\eqref{eq:loop_def4} a propagator
\begin{equation}
\frac{1}{p'^0+\omega_K(\vec{q}~)+\omega_{K^*}(\vec{p}~'+\vec{q}~)-i\varepsilon}\, 
\end{equation}
instead of the term in Eq.~\eqref{eq:loop_def4}
\begin{equation}
\frac{1}{P^0-\omega_D(\vec{p}~'+\vec{q}~)-\omega_{K^*}(\vec{p}~'+\vec{q}~)+i\varepsilon}\, .
\end{equation}
We find numerically that the corrections from inclusion of the new term are of the order of 10\% as an average. These uncertainties have to be assumed in our result, but are still small compared to the uncertainties in Eq.~\eqref{eq:widthresult} from other sources.
}

\section{Discussion}\label{Sec:dis}

{While in the refereeing process of the paper, a work appeared~\cite{Yang:2025dcg} evaluating the $D_{s1}(2460)$ and $D_{s1}(2536)$ decay to $D_s^+\pi^+\pi^-$. This work is timely and illustrating, and allows us to give a new perspective to our work.}

{Contrary to our work, that assume the $D_{s1}(2536)$ state to be mostly molecular from the $K^*D$ component, Ref.~\cite{Yang:2025dcg} considers the state mostly of $q\bar{q}$ type, based on the findings of Ref.~\cite{Yang:2021tvc}. The molecular components are $K^*D$ in $S$- and $D$-wave and barely amount to about 2\%. Interestingly, it is observed in \cite{Yang:2025dcg} that the $c\bar{s}(P)$ quark model component of the $D_{s1}$ state cannot decay to $c\bar{s}(S)+f_0(500)\to\pi\pi$, that is OZI forbidden, which forces the $D_{s1}\to D_s^+\pi^+\pi^-$ decay through a loop mechanism involving the molecular components, even if they are small.}

{Since the molecular components for $D_{s1}(2536)$ assumed in \cite{Yang:2025dcg} are different from those considered in the present work, it is not surprising that the line shapes of the $\pi\pi$ and $D_s\pi$ mass distributions are quite different in the two pictures, which is a welcome feature, because it means that the experimental implementation of the reaction can tell us about the basic molecular components present in the $D_{s1}(2536)$ state. This should be an incentive to measure the reaction.}

{From our point of view, the presence of the $K^*D$ component in the $D_{s1}(2536)$ is necessary, although we cannot quantify it. This statement is based on the observation that the $D_{s1}(2460)$ contains a large $D^*K$ molecular component, $(57\pm21\pm6)\%$ according to \cite{MartinezTorres:2014kpc} based on lattice QCD data, compatible with that found in \cite{Yang:2021tvc} of $(1-(52.4~^{+5.1}_{-3.8}))\%$. The $D^*K$ component is obtained because the $D^*K$ interaction is found attractive and strong enough to bind by itself a $D^*K$ molecule \cite{Lin:2024hys}. The $K^*D$ system also has a strong attractive interaction. Indeed, using the local hidden gauge approach with the exchange of vector mesons, the $D^*K$ interaction at threshold goes as $2m_{D^*}2m_K$ and the $K^*D$ interaction as $2m_{D}2m_{K^*}$ with the same coefficient. This means that the $K^*D$ interaction has a strength 1.7 times bigger than that of the $D^*K$ and should generate some important $K^*D$ component in a state orthogonal to the $D_{s1}(2460)$ with the same quantum numbers, in this case the $D_{s1}(2536)$. It is true that the binding of the $D_{s1}(2536)$ is about 220 MeV below the $K^*D$ threshold, while that of the $D_{s1}(2460)$ is only about 40 MeV below the $D^*K$ threshold, but given the much larger strength of the $K^*D$ interaction, as pointed above, it is natural to expect a much larger binding. Such large bindings are not absent in hadron molecules, and for instance the $f_2(1270)$ resonance appears as a bound state of two $\rho$ mesons in the work of \cite{Molina:2008jw,Geng:2008gx}. From the discussion above, it seems natural to expect a sizeable $K^*D$ component in the $D_{s1}(2536)$ state, and this has been our starting point. This idea might be in conflict with the zero $K^*D$ component of \cite{Yang:2021tvc}, but we think there is no conflict since in the analysis of the lattice QCD data done in \cite{Yang:2021tvc} the interpolating $K^*D$ fields were not considered, hence, a solution is found without the $K^*D$ component.}

{From our point of view there should be a sizeable $K^*D$ component in the $D_{s1}(2536)$ state, but experiments like the one studied here and others will tell us in the future. However, we should also not expect a 100\% molecular state as we have assumed so far. The $D_{s1}(2460)$ has about 60\% $D^*K$ component and we should not expect more than that. However, in the present calculation we have assumed the whole $D_{s1}(2536)$ to be of molecular nature, and this provided the coupling $g_{D_{s1}(2536),K^{*+}D^0}$ extracted from Ref.~\cite{Lin:2024hys} under this assumption. Should the molecular $K^*D$ component be smaller than 100\% this coupling would be smaller, and the probability would roughly be proportional to this coupling squared \cite{Gamermann:2009uq}. This means that the measurement of the total width of $D_{s1}(2536) \to D_s^+ \pi^+ \pi^-$ would serve as a measure of this component. If this experimental rate would be about 1/2 of our predictions and the shape of the distributions the same that we predict, we could come to the conclusion that the $K^*D$ molecular component is about 50\%. After the former discussion, it is clear that future measurements of the $D_{s1}(2536) \to D_s^+ \pi^+ \pi^-$ reaction, with precise mass distributions and integrated width should provided an excellent information to learn about the molecular components of the  $D_{s1}(2536)$ and their strength.}

\par
\vspace{1.5em}
\section{Summary} 
\label{Sec:conc}

In summary, our study highlights 
the relevance of triangle loop 
mechanisms in the $\dsd$ decay 
process, where intermediate meson-meson 
interactions play a crucial role in 
shaping the final-state mass distributions. 
Within our model, we assume that the 
decay proceeds via such loops, adopting 
a molecular picture for the axial meson 
$D_{s1}(2536)$, where this state is 
dynamically generated through the $K^*D$ 
interaction.

We have also emphasized the impact of 
dynamical effects driven by the proximity 
of intermediate thresholds. In the present 
study, these effects are particularly 
significant, giving rise to non-trivial 
structures in the spectra. Specifically, 
in the invariant mass distribution of the 
$ D^+_s\pi^+$ pair, the manifestation of 
threshold singularities becomes evident 
at values that coincide with the on-shell 
conditions of the intermediate $D^0 K^0$ 
and $D^+ K^+$ states.

{We have provided the mass distributions and the total decay width in absolute values, within a range of uncertainty. The calculations are based on the assumption of the $D_{s1}(2536)$ as a pure molecular $K^*D$ state. Interestingly, the $q\bar{q}$ components that one also expects for this state do not contribute to this process since it is OZI forbidden as noted in \cite{Yang:2025dcg}, hence the shape of the distribution is tied to the molecular components. Yet, the coupling of $D_{s1}(2536)$ to $K^*D$ would be smaller if there is a sizeable $q\bar{q}$ component in the state, which would revert into a smaller integrated width than the one that we have calculated. This is telling us that the actual measurement of the $D_{s1}(2536) \to D^+_s \pi^+ \pi^-$ mass distributions and decay rate can tell us much about the nature of the $D_{s1}(2536)$.}

Overall, these findings offer further 
insight into the role of hadronic loops 
and final-state interactions in shaping 
observable features—such as cusps and 
enhancements—in multi-body decay processes 
of heavy mesons.

\vspace{15pt}

\begin{acknowledgements}

This work is supported by the Spanish Ministerio de Economia y
Competitividad (MINECO) and European FEDER funds
under Contracts No. FIS2017-84038-C2-1-P B, PID2020-112777GB-I00, and
by Generalitat Valenciana under contract
PROMETEO/2020/023. This project has received funding from the European
Union Horizon 2020 research and innovation
programme under the program H2020-INFRAIA-2018-1, grant agreement No.
824093 of the STRONG-2020 project. This
work is supported by the Spanish Ministerio de Ciencia e Innovacion
(MICINN) under contracts PID2020-112777GB-I00,
PID2023-147458NB-C21 and CEX2023-001292-S; by Generalitat Valenciana
under contracts PROMETEO/2020/023 and
CIPROM/2023/59. Yi-Yao Li is supported in part by the Guangdong Provincial international exchange program for outstanding young talents of scientific research in 2024.
\end{acknowledgements}

\bibliography{Ds1_refs.bib}

\end{document}